\documentclass[aps,pre,twocolumn,groupedaddress,showpacs]{revtex4-1}
\newcommand{\bec}[1]{\mbox{\boldmath $ #1$}}
\usepackage{graphicx}
\usepackage{bm}
\usepackage{amssymb,amsfonts,amsmath}%
\usepackage{color}

\begin{document}
\title{Internal gravity waves in the energy and flux budget turbulence-closure theory
for shear-free stably stratified flows}
\author{N. Kleeorin$^{1,2,3}$}
%\email{nat@bgu.ac.il}
\author{I.~Rogachevskii$^{1,2,3}$}
\email{gary@bgu.ac.il}
\homepage{http://www.bgu.ac.il/~gary}
\author{I. A. Soustova$^{4}$}
%\email{soustova@appl.sci-nnov.ru}
\author{Yu.~I.~Troitskaya$^{4}$}
%\email{yuliya@hydro.appl.sci-nnov.ru}
\author{O. S. Ermakova$^{4}$}
%\email{o.s.ermakova@mail.ru}
\author{S. Zilitinkevich$^{3,5,6,7}$}
%\email{sergej.zilitinkevich@fmi.fi}

\bigskip
\affiliation{
$^1$Department of Mechanical Engineering, Ben-Gurion University of the Negev, P. O. B. 653, Beer-Sheva
 8410530, Israel
 \\
$^2$Nordita, Stockholm University and KTH Royal Institute of Technology, 10691 Stockholm, Sweden
 \\
$^3$Institute for Atmospheric and Earth System Research, University of Helsinki, 00014 Helsinki, Finland
 \\
$^4$Institute of Applied Physics of the Russian Academy of Sciences, 603950 Nizhny Novgorod, Russia
 \\
$^5$Finnish Meteorological Institute, 00101 Helsinki, Finland
 \\
$^6$Faculty of Geography,~Moscow State University, 119234 Moscow, Russia
\\
$^7$Tyumen State University, 625003 Tyumen, Russia
}

\date{\today}
\begin{abstract}
We have advanced the energy and flux budget (EFB) turbulence closure
theory that takes into account a two-way coupling between internal
gravity waves (IGW) and the shear-free stably stratified turbulence.
This theory is based on the budget equation for the total (kinetic plus
potential) energy of IGW, the budget equations for the kinetic and
potential energies of fluid turbulence, and turbulent fluxes of potential temperature
for waves and fluid flow. The waves emitted at a certain level, propagate
upward, and the losses of wave energy cause the production of turbulence energy.
We demonstrate that due to the nonlinear effects more intensive waves produce
more strong turbulence, and this, in turns, results in strong damping of IGW.
As a result, the penetration length of more intensive waves is shorter than
that of less intensive IGW. The anisotropy of the turbulence produced by less
intensive IGW is stronger than that caused by more intensive waves.
The low amplitude IGW produce turbulence consisting up to 90 \% of
turbulent potential energy. This resembles the properties of the observed high
altitude tropospheric strongly anisotropic (nearly two-dimensional) turbulence.
\end{abstract}

%e-print: NORDITA-2018-129

\maketitle
\section{Introduction}

The classical theory of atmospheric flows is based on seminal papers by Rayleigh, Richardson, Prandtl, Kolmogorov, Obukhov and Monin (see, e.g., Refs.~\cite{MY71,MY75}). This theory implies that any turbulent flow can be considered as a superposition of the fully organised mean flow, and the fully chaotic turbulence characterised by the forward energy cascade from the larger eddies to smaller, resulting in the viscous energy-dissipation at the smallest eddies, with a pronounced inertial interval in the energy spectrum that is fully determined by the energy-dissipation rate \cite{K41}. The local characteristics of turbulence, in particular, turbulent fluxes that appear in the mean-flow equations, are generally controlled by the local features of the mean flow. It is also assumed that the turbulent flux of any quantity can be expressed as a product of the mean gradient of the quantity multiplied by a turbulent-exchange coefficient. This concept of down-gradient transport reduces the closure problem to determination of the turbulent-exchange coefficients, usually taken proportional to the turbulent kinetic energy and timescale (see, e.g., Refs.~\cite{MY71,MY75}). This has been formulated for neutrally stratified flows.

However, atmospheric flows often include, besides Kolmogorov turbulence, another type of motions, associated with the development of large-scale structures, e.g, large-scale coherent semi-organized structures (i.e., cloud cells and cloud streets) in turbulent convection \cite{Z91,EB93,AZ96,Z02,EKRZ02} and internal gravity waves in stably stratified turbulence \cite{B74,GH75,M01,N02,BU09,S10}.
The majority of efforts in development of the turbulence closure models for meteorological applications over half a century have been limited to ''mechanical closures" based on the sole use of the turbulent kinetic energy equation, disregarding turbulent potential energy, and daring only cautious corrections to the concept of down-gradient transport.

In stable stratification, such closures have resulted in the erroneous conclusion that shear-generated turbulence inevitably decays and that the flow becomes laminar in ''supercritical" stratifications (at Richardson numbers exceeding some critical value, see, e.g., \cite{C61,M86}). Obvious contradictions of this conclusion via the well-documented universal presence of turbulence in strongly “supercritical” conditions typical of the free atmosphere and the deep ocean (see, e.g., Refs.~\cite{SF01,O01,BAN02,PAR02,MO02,LA04,M14}) were attributed to some unknown mechanisms and, in practical applications, mastered heuristically (see overviews in Refs.~\cite{M14,CAN09}). The decay of strongly stably stratified turbulence in direct numerical simulations (DNS) has been explained by the effect of diminishing the effective Reynolds number, which comes into play in the not-high-Reynolds-number flows in DNS, but remains insignificant in the very-high-Reynolds-number atmospheric flows.

These principal problems call for a revision of the traditional theory of atmospheric turbulence. The strongest motivation for the revision comes from the need to improve the modern numerical weather prediction, air quality, and climate models, in which turbulent planetary boundary layers couple the atmosphere with underlying land/water/ice surfaces. Stably stratified turbulence determines the vertical turbulent transport of energy and momentum and the turbulent diffusion of pollutants, aerosols, and other admixtures in the free atmosphere.

Numerous alternative turbulence closures in stratified turbulence have been formulated
using the budget equations for various turbulent parameters (in addition to the turbulent
kinetic energy) together with heuristic hypotheses and
empirical relationships (see reviews \cite{WT03,UB05}).
Two-point turbulent closures have been developed as well (see reviews \cite{C01,SC18}),
which take into account a very detailed scale-by-scale and directional anisotropy,
that is almost lost in single-point closures.

Key ingredients of stably stratified turbulence are internal gravity waves.
In atmospheric and oceanic turbulence they have been a subject of intense research
(see, e.g., \cite{GM79,IOS83,OT87,HM06,JDS06,S07,PZ14,SN15,YM15}).
In the atmosphere, internal gravity waves exist
at scales ranging from meters to kilometers, and are measured by direct probing
or remote sensing using radars and lidars.
The sources of internal gravity waves can be flows over complex terrain, strong wind
shears, convective and other local-scale motions underlying the stably stratified
layer, and wave-wave interactions \cite{C99,SS02,FA03}.

The different nature of fluctuations caused by turbulence and waves in stratified flows has been pointed out in
\cite{JRF05}.
The role of waves in turbulence closure models has been discussed in \cite{JSG03,BP04,ST04}
An additional negative term in the TKE budget equation (the rate of transfer of TKE into potential energy of wave-like motions) has been included in \cite{BP04}. It was noted that with increasing stability, the wave-like motions dominate in comparison with velocity and buoyancy fluctuations of stratified turbulence, and fluctuations caused by waves suppress vertical mixing (see also \cite{UB05}).

Analysis of the budgets of the wave kinetic energy and the wave temperature variance has been done in \cite{FE81,FEF84,EF84,F88,EF93}.
They found significant buoyant production of the wave energy despite the strong static stability and energy transfer from waves to turbulence.
Different aspects related to the effects of
internal gravity waves (IGW) have been also discussed in \cite{SC18}, where
it has been stressed that in the limit of small Froude number,
IGW only affect a poloidal part of the flow.
A strong toroidal cascade coexisting with weak IGW cascade
has been found as well \cite{SC18}.

Numerous high resolution DNS of stably stratified turbulence
with IGW have been performed as well (see, e.g., \cite{LI06,CM15,PR18,DOZ13,DO15}).
The role of IGW and their interaction with the
large-scale flow of vertically sheared horizontal winds has been studied in \cite{CM15}.
It has been shown that most of the energy is concentrated along a
dispersion relation that is Doppler shifted by the horizontal winds. They pointed out that when uniform winds
are let to develop in each horizontal layer of the flow, waves whose phase velocity is equal to the horizontal
wind speed have negligible energy, which indicates a nonlocal transfer of their energy to the mean flow.
Scaling laws for mixing and dissipation in unforced stratified turbulence have been found in \cite{PR18}.
Three regimes characterised by Froude number, namely (i) dominant waves, (ii) eddy-wave
interactions and (iii) strong turbulence have been observed in \cite{PR18}.
An interaction between large-scale IGW and turbulent layer above
the pycnocline (where the density gradient is largest)
has been studied using DNS in \cite{DOZ13,DO15}.
They have demonstrated that in the absence of IGW, the turbulence decays
and most of the turbulent energy is concentrated at the pycnocline center.
The turbulent eddies are collapsed
in the vertical direction and acquire the “pancake” shape.
The internal gravity waves significantly increase turbulent energy
\cite{DOZ13,DO15}.

The energy- and flux-budget (EFB) theory of turbulence closure for stably stratified dry atmospheric flows has been recently developed in Refs.~\cite{ZKR07,ZKR08,ZKR09,ZKR10,ZKR13}. In accordance with wide experimental evidence, the EFB theory shows that high-Reynolds-number turbulence is maintained by shear in any stratification, and the ''critical Richardson number", treated over decades as a threshold between the turbulent and laminar regimes, factually separates two turbulent regimes: the strong turbulence, typical of atmospheric boundary layers, and the weak three-dimensional turbulence, typical of the free atmosphere or deep ocean and characterized by sharp decrease in heat transfer in comparison to momentum transfer. The principal aspects of the EFB theory have been verified against scarce data from the atmospheric experiments, direct numerical simulations, large-eddy simulations (LES) and laboratory experiments relevant to the steady state turbulence regime.

In stably stratified turbulence, large-scale internal gravity waves cause additional vertical turbulent flux of momentum and additional productions of turbulent kinetic energy (TKE), turbulent potential energy (TPE) and turbulent flux of potential temperature \cite{ZKR09}. For the stationary, homogeneous regime, the EFB theory in the absence of the large-scale IGW yields universal dependencies of the flux Richardson number, the turbulent Prandtl number, the ratio of TKE to TPE, and the normalised vertical turbulent fluxes of momentum and heat on the gradient Richardson number, Ri (see Refs.~\cite{ZKR07,ZKR13}). Due to the large-scale IGW, these dependencies lose their universality.  The maximal value of the flux Richardson number (universal constant  0.2-0.25 in the no-IGW regime) becomes strongly variable in the turbulence with large-scale IGW. In the vertically homogeneous stratification, the flux Richardson number increases with increasing wave energy and can even exceed 1.
The large-scale internal gravity waves also reduce anisotropy of turbulence. Indeed, in contrast to the mean wind shear, which generates only horizontal component of the turbulent kinetic energy, IGW generate both horizontal and vertical components of TKE. These waves increase the share of TPE in the turbulent total energy (TTE = TKE + TPE). A well-known effect of IGW is their direct contribution to the vertical transport of momentum. Depending on the direction (downward or upward), IGW either strengthen or weaken the total vertical flux of momentum. Predictions from this theory \cite{ZKR09} are consistent with available data from atmospheric and laboratory experiments, DNS and LES.

In the theory discussed in \cite{ZKR09},
the stably stratified turbulence is produced
by a large-scale wind shear. This theory takes only
into account a one-way coupling corresponding to
the effect of large-scale IGW with random phases
on stably stratified turbulence,
while the feedback effect of the
turbulence on IGW has been ignored in \cite{ZKR09}.
In view of applications, the theory discussed in \cite{ZKR09} describes only
turbulence in the lower troposphere (up to the altitudes about 1 -- 1.5 km).

The goal of the present study is to investigate the two-way coupling between large-scale IGW and shear-free stably stratified turbulence. This implies that the turbulence is produced solely by dissipation of IGW propagating in stratified turbulent flows.
In the analysis, we use the budget equations for the kinetic and
potential energies of both, fluid turbulence and large-scale IGW with random phases.
We also apply the budget equations for turbulent heat flux and momentum.
We demonstrate that due to the nonlinear effects the penetration length of the more intense IGW is less than that for the less intensive IGW (with lower wave energy). The low amplitude IGW produce turbulence consisting up to 90 \% of potential energy.
The results of the present study describe only
the upper troposphere (located at the altitudes about 10 -- 15 km),
see, e.g., \cite{PZ14,SN15,YM15}, and references therein.

This paper is organized as follows.
In Section II we outline properties of the internal gravity waves
propagating in a fluid in the absence of turbulence. We also discuss here
the energy budget equations for IGW.
In Section III we formulate governing equations for the energy and flux budget
turbulence-closure theory for stably stratified turbulence with large-scale IGW.
In Section IV we study the effects of large-scale IGW on turbulence for the steady-state
regime.
The two-way coupling between turbulence and large-scale IGW is analysed in Section V.
Finally, conclusions are drawn in Section VI.

\section{Large-scale IGW in the stably stratified flows}

In this study we focus on the effect of large-scale internal
gravity waves on the stably stratified turbulence.

\subsection{Linear IGW in the absence of turbulence}

First we outline properties of the internal
gravity waves propagating in a fluid in the absence of turbulence and neglecting dissipations.
These waves are described by the following equations:
\begin{eqnarray}
{\partial \tilde {\bm V}^{\rm W} \over \partial t} + (\tilde{\bm V}^{\rm W} \cdot {\bm \nabla}) \tilde{\bm V}^{\rm W} &=& - {{\bm \nabla} \tilde P^{\rm W} \over \rho_0} + \beta \, \tilde \Theta^{\rm W} {\bm e} ,
 \label{B1}\\
{\partial \tilde \Theta^{\rm W} \over \partial t} + (\tilde{\bm V}^{\rm W} \cdot {\bm \nabla}) \tilde \Theta^{\rm W} &=&  - \beta^{-1} N^2 \, \tilde V^{\rm W}_j e_j,
 \label{B2}
\end{eqnarray}
where $\tilde {\bm V}^{\rm W}$, $\tilde \Theta^{\rm W}$  and $\tilde P^{\rm W}$ are the velocity, potential temperature and pressure characterizing IGW, ${\bm e}$ is the vertical unit vector, $\beta=g/T_0$ is the buoyancy parameter,  ${\bm g}$ is the acceleration due to gravity,  $N= (\beta |\nabla_z \Theta|)^{1/2}$ is the Brunt-V\"{a}is\"{a}l\"{a} frequency, $\Theta$ is the potential temperature of fluid defined as $\Theta= T \,(P_0/ P)^{1-1/\gamma}$, where $T$ is the absolute temperature, $T_0$  is its reference value, $P$ is the fluid pressure, $P_0$  is its reference value,  $\gamma=c_p/c_v =1.41$ is the specific heats ratio, and $\rho_0$ is the fluid density. The potential temperature $\tilde \Theta^{\rm W}$ for waves is defined in a similar way.
Equations~(\ref{B1}) and~(\ref{B2}) are written in the Boussinesq approximation with the continuity equation,
div $\tilde {\bm V}^{\rm W} =0$.

Solution of linearized equations~(\ref{B1}) and~(\ref{B2}) is given by
\begin{eqnarray}
\tilde {\bm V}^{\rm W} &=& \left({\bm e} - {{\bm k}_h k_z \over k_h^2} \right) \, V_\ast^{\rm W}(z) \, \cos \left[\varphi({\bm r}) - \omega t\right],
 \label{B3}\\
\tilde \Theta^{\rm W} &=& {N^2(z) \over \omega \beta} \, V_\ast^{\rm W}(z) \, \sin \left[\varphi({\bm r}) - \omega t\right],
 \label{B4}\\
 \tilde P^{\rm W} &=& - \rho_0 \, \left({k_z \, \omega  \over k_h^2} \right) \, V_\ast^{\rm W}(z) \, \cos \left[\varphi({\bm r}) - \omega t\right],
 \label{B5}
\end{eqnarray}
(see, e.g., \cite{M01,N02,T73}), where  $\varphi({\bm r})$ is the wave phase, ${\bm k}={\bm k}_h + {\bm e} k_z$  is the wave vector; ${\bm k}_h = (k_x, k_y)$  is the horizontal wave vector,
$V_\ast^{\rm W}(z)$ is the wave velocity amplitude,
and the frequency of IGW is given by
\begin{eqnarray}
\omega = N(z) {k_h \over k} .
 \label{GGG15}
\end{eqnarray}
Propagation of IGW in an inhomogeneous medium is determined in the approximation of geometrical optics by the following Hamiltonian equations:
\begin{eqnarray}
{\partial {\bm r} \over \partial t} = {\partial \omega \over \partial {\bm k}},
\label{BBB6}\\
{\partial {\bm k} \over \partial t} = - {\partial \omega \over \partial {\bm r}},
\label{B6}
\end{eqnarray}
(see, e.g., \cite{W62}), where ${\bm r}$ is the radius-vector of the center of the wave packet. Since the Brunt-V\"{a}is\"{a}l\"{a} frequency $N=N(z)$ depends on the vertical coordinate, Eq.~(\ref{B6}) yields ${\bm k}_h=$ const and the vertical component of the wave vector depends on $z$, i.e., $k_z=k_z(z)$.  Equations~(\ref{GGG15}) and~(\ref{B6}) for the IGW with a fixed frequency allow us to determine $k(z)$ as
\begin{eqnarray}
k(z)=k_0 \, {N(z) \over N(z_0)},
\label{B7}
\end{eqnarray}
where $z=z_0$  is the hight where the IGW source is located and $k_0=k(z=z_0)$.
Equation~(\ref{B7}) determining the $z$-dependence of $k(z)=(k_z^2 + k_h^2)^{1/2}$  implies that the vertical wave numbers, $k_z(z)$, change when the IGW propagates through the stably stratified flow.
When the IGW source is located at the surface (for $z_0=0$), the vertical wave number is
$k_z(z)=k_h [N^2(z)/\omega^2 - 1]^{1/2}$. When the IGW source is located at the upper boundary of the layer under consideration (for $z_0=H$), the vertical wave number is given by $k_z(z)=-k_h [N^2(z)/\omega^2 - 1]^{1/2}$.
It follows from these expressions that $k(z) = k_h N(z)/\omega$.

\subsection{Budget equation for total energy of IGW}

Let us study the effect of large-scale IGW on the stably stratified turbulence.
We assume that the wavelengthes and periods of the large-scale IGW are
much larger than the turbulence spatial scales and timescales, respectively.
This allows us to treat the large-scale IGW with respect to turbulence
as a kind of mean flows with random phases.
We neglect small molecular dissipation of IGW for large Reynolds and P\'{e}clet numbers.
At the low frequency part of the IGW spectra, we limit our analysis to frequencies
essentially exceeding the Coriolis frequency.

We consider the flow fields as a superposition of three components: mean fluid velocity,
${\bm U}(t,z)$, fluctuations of fluid velocity, ${\bm u}(t,{\bm r})$, and random large-scale IGW velocity field,  ${\bm V}^{\rm W}(t,{\bm r})$, i.e., the total velocity is ${\bm v}={\bm U} + {\bm u} + {\bm V}^{\rm W}$.
To determine the random large-scale wave velocity field,  ${\bm V}^{\rm W}(t,{\bm r}) \equiv \langle {\bm v} \rangle - \overline{\langle {\bm v} \rangle}$,
besides the ensemble averaging over turbulent fluctuations (denoted by the angular brackets), we also perform an averaging in time over the IGW period (denoted by overbar).
We also use similar decompositions for the total potential temperature,  $\Theta_{\rm tot} =\Theta + \theta + \Theta^W$, and for the total pressure,  $P_{\rm tot}=P + p + P^{\rm W}$, where
$P$ and $\Theta$ are the mean potential temperature and pressure, respectively,
$p$ and $\theta$ are fluctuations of the potential temperature and pressure, respectively, and
the wave fields are determined as $\Theta^{\rm W}(t,{\bm r}) = \langle \Theta_{\rm tot} \rangle - \overline{\langle \Theta_{\rm tot} \rangle}$ and $P^{\rm W}(t,{\bm r}) = \langle P_{\rm tot} \rangle - \overline{\langle P_{\rm tot} \rangle}$.
We assume that the mean fields depend on $z$-coordinate and time, while the large-scale wave fields depend on all three coordinates and can be represented as an ensemble of wave packets with narrow frequency range and random phases.

In this study we consider a shear-free turbulence, and for simplicity we assume that the mean fluid velocity is zero. We consider low-amplitude approximation for the large-scale IGW and neglect the wave-wave interactions, but take into account the interactions between turbulence and the large-scale IGW.
Equations for the large-scale IGW in stably stratified turbulence are given by
\begin{eqnarray}
{\partial V^{\rm W}_i \over \partial t} &=& - {\nabla_i P^{\rm W} \over \rho_0} + \beta \, \Theta^{\rm W} e_i - \nabla_j \tau_{ij}^{\rm W},
 \label{B8}\\
{\partial \Theta^{\rm W} \over \partial t} &=& - \beta^{-1} N^2 \, V^{\rm W}_j e_j - \nabla_j F_{j}^{\rm W} ,
 \label{B9}
\end{eqnarray}
where the interaction between the turbulence and the large-scale IGW is described by the effective Reynolds stress tensor, $\tau_{ij}^{\rm W} = \langle v_i v_j \rangle - \overline{\langle v_i v_j \rangle}$,
and the effective flux of potential temperature, $F_{i}^{\rm W} = \langle v_i \Theta_{\rm tot} \rangle - \overline{\langle v_i \Theta_{\rm tot} \rangle}$.
Equations~(\ref{B8})--(\ref{B9}) are mean-field equations,
where the divergence of the Reynolds stress
in the mean momentum equation, and the divergence of the turbulent
heat flux in the mean potential temperature equation
determine effects of turbulence on the mean fields.
To derive Eqs.~(\ref{B8})--(\ref{B9}) for IGW, we obtain two system of equations:
the first system is obtained by the ensemble averaging of the exact momentum
and potential temperature equations over turbulent fluctuations (denoted by the angular brackets);
the second system is obtained by the additional time averaging of the first system of
mean-field equations over the IGW periods (denoted by overbar).
Equations~(\ref{B8})--(\ref{B9}) are obtained by the subtraction of the second system
of equations from the first system.

Using Eqs.~(\ref{B8})--(\ref{B9}), we derive budget equation for the total wave energy,
$E^{\rm W}=E_K^{\rm W}+E_P^{\rm W}$,
\begin{eqnarray}
{\partial E^{\rm W} \over \partial t} + {\bm \nabla} \cdot {\bm \Phi}^{\rm W} = -D^{\rm W} ,
 \label{B10}
\end{eqnarray}
where $E_K^{\rm W}=\overline{({\bm V}^{\rm W})^2}/2$ is the turbulent kinetic wave energy, $E_P^{\rm W}=(\beta/N)^2 \, \overline{(\Theta^{\rm W})^2}/2$ is the turbulent potential wave energy, ${\bm \Phi}^{\rm W} = \rho_0^{-1} \overline{{\bm F}^{\rm W} \, P^{\rm W}}$ is the flux of the total wave energy, $E^{\rm W}$, and
$D^{\rm W}$ is the dissipation rate of large-scale IGW given by
\begin{eqnarray}
D^{\rm W} &=&  - {\beta^2 \over N^2}
\left(\overline{F_{j}^{\rm W} \, \nabla_j \Theta^{\rm W}} - \overline{F_{j}^{\rm W} \, \Theta^{\rm W}} \, \nabla_j \ln N^2 \right)
\nonumber\\
&& - \overline{\tau_{ij}^{\rm W} \, \nabla_j V^{\rm W}_i},
 \label{B11}
\end{eqnarray}
is the dissipation rate of the large-scale IGW that is determined by the work of turbulence caused by the interaction with the large-scale IGW. In the gradient approximation, the dissipation rate of the large-scale IGW is positive, $D^{\rm W}>0$. Indeed, as has been shown in \cite{M01} (see Eq.~(3.72) in \cite{M01}), the flux ${\bm \Phi}^{\rm W}$ of the total wave energy, $E^{\rm W}$, is given by ${\bm \Phi}^{\rm W} ={\bm V}^{\rm g} E^{\rm W}$, where ${\bm V}^{\rm g}=\pm \omega \,({\bm k}/k^2 - {\bm k}_h/k_h^2)$ is the group velocity of the large-scale IGW.

Let us obtain a steady-state solution of budget equation~(\ref{B10})
for the total wave energy considering two cases:

(i) {\it Non-dissipative large-scale IGW}, i.e., the dissipation rate of the total wave energy vanishes, $D^{\rm W}=0$. In this case steady-state solution of Eq.~(\ref{B10})
reads: ${\bm \nabla}~\cdot~({\bm V}^{\rm g} E^{\rm W})=0$.
Since ${\bm k}_h=$ const and $\omega=$ const, we find that the vertical profile of the wave amplitude in the solution for $\tilde {\bm V}^{\rm W}(t,{\bm r})$ [see Eq.~(\ref{B3})] is given by:
\begin{eqnarray}
|V_\ast^{\rm W}(z)|=|V_0| \, [N^2(z)/\omega^2 - 1]^{-1/4} ,
\label{B12}
\end{eqnarray}
where $V_0$ is the constant of integration.
This profile is in agreement with that obtained in \cite{M01}.

(ii) {\it Dissipative large-scale IGW}.
In this case the dissipation rate of the total wave energy is
\begin{eqnarray}
D^{\rm W}= K_H(1 + {\rm Pr}_{_{T}}) k^2 E^{\rm W} ,
\label{L1}
\end{eqnarray}
which is derived for a narrow wave packet, and
a homogeneous and isotropic turbulence (see Appendix~\ref{Appendix-A}), where
${\rm Pr}_{_{T}}=K_M/K_H$ is the turbulent Prandtl number, $K_M$ is the eddy viscosity
and $K_H$ is the eddy diffusivity.
Thus, the steady-state solution of Eq.~(\ref{B10})
reads: ${\bm \nabla}~\cdot~({\bm V}^{\rm g} E^{\rm W})=-D^{\rm W}$, which can be rewritten as
\begin{eqnarray}
{d \Phi^{\rm W}_z \over dz} &=& - \sigma_g \Phi^{\rm W}_z ,
\label{B13}
\end{eqnarray}
where $\Phi^{\rm W}_z= V^{\rm g}_z E^{\rm W}$ and
\begin{eqnarray}
\sigma_g &=& K_H(1 + {\rm Pr}_{_{T}}) k_h^3 \left(N^2 - \omega^2\right)^{-1/2} \left({N \over \omega}
\right)^4 .
\label{B14}
\end{eqnarray}
Equation~(\ref{B13}) yields
\begin{eqnarray}
|V_\ast^{\rm W}(z)|=|V_0| \, |N^2(z)/\omega^2 - 1|^{-1/4} \exp
\left(- {\tau_g \over 2}\right) ,
\nonumber\\
\label{B15}
\end{eqnarray}
where $\tau_g=\int_0^z \sigma_g(z') \,dz'$. In the absence of turbulence ($K_M, K_H \to 0$), the parameter $\tau_g$ tends to 0, so that Eq.~(\ref{B15}) coincides with
Eq.~(\ref{B12}). It follows from Eq.~(\ref{B15}) that in the absence of turbulence in the vicinity of a resonance, $\omega =N(z=z_r)$, the amplitude of the large-scale IGW tends to infinity. This implies that the low amplitude approximation does not valid, and the nonlinear effects (e.g., the wave braking) should be taken into account. For instance, the wave braking can cause an additional production of turbulence. On the other hand, in the presence of turbulence the infinite growth of the wave amplitude does not occur if the first and the second spatial derivatives of the Brunt-V\"{a}is\"{a}l\"{a} frequency vanish at the surface $z=z_r$ \cite{M01}.

\section{Energy and flux budget (EFB) model for turbulence with large-scale IGW}

In this section we formulate the budget equations for stably stratified turbulence
with large-scale IGW.

\subsection{Budget equations for turbulence with large-scale IGW}

In the framework of energy and flux budget turbulence  model, we use
the budget equations for the turbulent kinetic energy (TKE), $E_K=\overline{\langle {\bm u}^2 \rangle}/2$, the intensity of the potential temperature fluctuations, $E_\theta=\overline{\langle \theta^2 \rangle}/2$, and the vertical turbulent flux, $F_z = \overline{\langle u_z \theta \rangle}$, of potential temperature
accounting for large-scale IGW:
\begin{eqnarray}
{DE_K \over Dt} + \nabla_z \, \Phi_{K} &=& \beta \, F_z - \varepsilon_K
- \overline{\tau_{ij}^{\rm W} \, \nabla_j V^{\rm W}_i} + \beta \, \overline{V^{\rm W}_z \Theta^{\rm W}},
\nonumber\\
 \label{C1}\\
{D E_\theta \over Dt} + \nabla_z \, \Phi_\theta &=& - F_z \, \nabla_z \Theta - \varepsilon_\theta - \overline{F_{j}^{\rm W} \, \nabla_j \Theta^{\rm W}},
 \label{C2}\\
{D F_z \over Dt} + \nabla_z \, \Phi_F &=& \beta \, \overline{\langle \theta^2
\rangle} - {1\over \rho_0} \, \overline{\langle \theta \,
\nabla_z p \rangle} - \overline{\langle u_z^2 \rangle} \,
{\nabla}_z \, \Theta
\nonumber\\
&& - \varepsilon_z^{(F)} - \overline{\tau_{j3}^{\rm W} \, \nabla_j \Theta^{\rm W}}
- \overline{F_{j}^{\rm W} \, \nabla_j V^{\rm W}_z},
\nonumber\\
 \label{C3}
\end{eqnarray}
(see, e.g., Ref. \cite{ZKR09}),
where $D / Dt = \partial /\partial t + {\bm U} {\bf \cdot} \bec{\nabla}$.
The terms $\Phi_{K}$, $\Phi_\theta$ and $\Phi_F$ include
the third-order moments. In particular, $\Phi_{K} = \rho_0^{-1} \overline{\langle u_z \, p\rangle} + (1/2) \overline{\langle u_z \, {\bf u}^2 \rangle}$ determines the flux of $E_K$;
$\, \Phi_\theta = \overline{\langle u_z \, \theta^2 \rangle}/2$ determines the flux of $E_\theta$ and $\Phi_F = \overline{\langle u_z^2 \theta\rangle} + \rho_0^{-1} \, \overline{\langle \theta \, p \rangle} / 2$ determines the flux of $F_z$.
The terms $\varepsilon_K=E_K/t_{T}$, $\varepsilon_\theta=E_\theta/(C_p \, t_{T})$ and
$\varepsilon^{(F)}=F_z/(C_F \, t_{T})$ are the dissipation rates of the turbulent kinetic energy $E_K$, the intensity of the potential temperature fluctuations $E_\theta$ and the vertical turbulent heat flux $F_z$. These  dissipation rates are expressed using the Kolmogorov  hypothesis \cite{K41}, where $t_{T}=\ell_0 /E_K^{1/2}$ is the turbulent dissipation timescale, $\ell_0$ is the integral scale of turbulence, and $C_p$  and $C_F$ are dimensionless constants. The turbulent potential energy (TPE), $E_P$, is defined as $E_P = \beta^2 E_\theta / N^2$.

The third term in the right hand side of Eq.~(\ref{C3}) contributes to the traditional
gradient turbulent flux (proportional to $- {\nabla}_z \, \Theta)$ of potential temperature,
while the first and the second terms in the right hand side
of Eq.~(\ref{C3}) describe a non-gradient contribution to the vertical turbulent flux of
potential temperature. In stably stratified flows
the gradient and non-gradient contributions to the vertical turbulent flux of
potential temperature have opposite signs. This implies that
the non-gradient contribution decreases the traditional gradient turbulent flux.

The budget equations for the turbulent kinetic energies,
$E_\alpha = \overline{\langle u_\alpha^2\rangle}/2$, along the $x$, $y$
and $z$ directions can be written as follows:
\begin{eqnarray}
{DE_\alpha \over Dt} + \nabla_z \, \Phi_{\alpha} &=& \delta_{\alpha 3} \, \beta \, F_z - \varepsilon_{\alpha\alpha}^{(\tau)} + {1 \over 2} Q_{\alpha\alpha}
- \overline{\tau_{\alpha j}^{\rm W} \, \nabla_j V^{\rm W}_\alpha} ,
\nonumber\\
 \label{C4}
\end{eqnarray}
where $\alpha=x,y,z$, the term $\varepsilon_{\alpha\alpha}^{(\tau)} = E_\alpha/3t_{T}$
is the dissipation rate of the turbulent kinetic energy components,
$\Phi_{\alpha}= \rho_0^{-1} \overline{\langle u_\alpha \, p\rangle} + (1/2) \overline{\langle u_z \, u_\alpha^2 \rangle}$ determines the flux of $E_\alpha$, the term $Q_{\alpha\alpha}$
is correlations between the fluctuations of the pressure, $p$, and the small-scale velocity shears: $Q_{ij} = \rho_0^{-1} (\overline{\langle p \nabla_i u_j\rangle} + \overline{\langle p \nabla_j u_i\rangle})$.
In Eq.~(\ref{C4}) we do not apply the summation convention for the double Greek indices.
Equations~(\ref{C1})--(\ref{C4}) are obtained by averaging over the ensemble of turbulent fluctuations and over the period of large-scale IGW. These equations without the large-scale IGW terms can be found in \cite{ZKR07,ZKR08,ZKR13} (see also \cite{KF94,CM93,CCH02,OT87}).

Equations~(\ref{C1})--(\ref{C4}) contain production terms caused
by the large-scale IGW (see \cite{ZKR09}).
In particular, the terms $- \overline{\tau_{ij}^{\rm W} \, \nabla_j V^{\rm W}_i}
+ \beta \, \overline{V^{\rm W}_z \Theta^{\rm W}}$ in the right hand side of Eq.~(\ref{C1})
determine the production rate of TKE, and the term $\overline{F_{j}^{\rm W} \, \nabla_j \Theta^{\rm W}}$
in Eq.~(\ref{C2}) contributes to the production rate of TPE, while the terms
$- \overline{\tau_{j3}^{\rm W} \, \nabla_j \Theta^{\rm W}}
- \overline{F_{j}^{\rm W} \, \nabla_j V^{\rm W}_z}$ in Eq.~(\ref{C3}) describe
the production rate of the vertical turbulent flux, $F_z$, of potential temperature.
To close the system of the budget equations~(\ref{C1})--(\ref{C4}), one needs to determine
the Reynolds stress for the wave fields, $\tau_{ij}^{\rm W}$, and the turbulent flux of potential temperature
for the wave fields, ${\bm F}^{\rm W}$, which have been derived in \cite{ZKR09}:
\begin{eqnarray}
\tau_{ij}^{\rm W} &=& - C_\tau \, t_T \left(\tau_{im} \nabla_m V^{\rm W}_j + \tau_{jm} \nabla_m V^{\rm W}_i \right) ,
 \label{C5}\\
F_{i}^{\rm W} &=& - C_F \, t_T \left(\tau_{im} \nabla_m \Theta^{\rm W} + \tau_{i3}^{\rm W} \nabla_z \Theta + F_{m} \nabla_m V^{\rm W}_i \right) .
\nonumber\\
 \label{C6}
\end{eqnarray}
These quantities are caused by interactions between the large-scale IGW and turbulence. They are determined by subtracting of the ensemble-averaged equations (but not averaged over the IGW period) from exact equations for these quantities, assuming that $\omega t_T \ll 1$.
To derive Eqs.~(\ref{C5}) and~(\ref{C6}), we
assume that the effective dissipation rate, $\varepsilon_{ij}^{(\tau,W)}$, of
the Reynolds stress for the wave fields, $\tau_{ij}^{\rm W}$, and
the effective dissipation rate, $\varepsilon_{i}^{(F,W)}$,
of the turbulent flux of potential temperature
for the wave fields, ${\bm F}^{\rm W}$, are expressed as
$\varepsilon_{ij}^{(\tau,W)} =
\tau_{ij}^{\rm W}/(C_\tau \, t_T)$ and $\varepsilon_{i}^{(F,W)} =
F_{i}^{\rm W}/(C_F \, t_T)$, respectively, where $C_\tau$ is a dimensionless constant.
In Eqs.~(\ref{C5}) and~(\ref{C6}) we also
omit the terms which are quadratic in wave amplitude,
because they do not contribute to the correlations
$\overline{\tau_{ij}^{\rm W} \, \nabla_j V^{\rm W}_i}$
and  $\overline{\tau_{ij}^{\rm W} \, \nabla_j \Theta^{\rm W}}$
entering in Eqs.~(\ref{C1})--(\ref{C4}) \cite{ZKR09} (see also \cite{ZKR13}).

If there is no separation of timescales between the
turbulent scales and the IGW scales, there are additional correction factors
in the right hand side of Eqs.~(\ref{C5}) and~(\ref{C6}). In particular,
the right hand side of Eq.~(\ref{C5}) is multiplied by the factor
$[1 + (C_\tau \omega t_{_{T}})^2]^{-1}$, while
the right hand side of Eq.~(\ref{C6}) is multiplied by the factor
$[1 + (C_F \omega t_{_{T}})^2]^{-1}$.
However, since the free constants $C_\tau$ and $C_F$ are small, i.e.,
$C_\tau^2 \approx C_F^2 \approx 10^{-2}$ (see Sect.~IV),
these correction factors are of the order of 1 even when there is
no separation of the timescales.
On the other hand, any mean-field theory formally requires
a separation of scales for its validity.

\subsection{Summary of assumptions and steps of derivations}

In our analysis, we use budget equations for the one-point second moments
for the following reasons.
We develop a mean-field theory and do not study small-scale structure
of turbulence. In particular, we study  large-scale long-term dynamics, i.e.,
we consider effects in the
spatial scales which are much larger than the integral scale of turbulence
and in timescales which are much longer than the turbulent timescales.
We limit our analysis to the geophysical flows, in which the vertical variations
of the mean fields are much larger than the horizontal variations, so that the
terms associated with the horizontal gradients in the budget equations for turbulent
statistics can be neglected. For instance, in typical atmospheric flows, the vertical
scales are much smaller than the horizontal scales, so that the mean vertical velocity
is much smaller than the horizontal velocity.

We have made the following assumptions related to internal gravity waves:
\begin{itemize}
\item{The periods and wavelengths of IGW are larger than the turbulent correlation time
and the integral scale of turbulence.
}
\item{We assume that the large-scale wave fields
can be represented as an ensemble of wave packets with narrow
frequency range and random phases.
}
\item{We restrict the analysis to a low-amplitude approximation for the large-scale IGW
and neglect the wave-wave interactions, but take into account the interactions
between turbulence and the large-scale IGW. We leave an
account for the wave-wave interactions for further study.
}
\item{The ensemble of the large-scale IGW has a power-law spectrum
and is isotropic in the horizontal plane.
}
\item{We neglect small molecular dissipation of IGW considering turbulence with
large Reynolds and P\'{e}clet numbers.
At the low frequency part of the IGW spectra, we limit our analysis to frequencies
essentially exceeding the Coriolis frequency.
}
\end{itemize}

We apply the energy and flux budget turbulence closure model, which
assumes the following:
\begin{itemize}
\item{The characteristic times of variations of
the turbulent kinetic energy, the turbulent potential energy,
and the vertical turbulent flux $F_z$ of potential temperature
are much larger than the turbulent timescale.
This allows us to obtain a steady-state solution
for the budget equations of TKE, TPE and $F_z$
for a homogeneous stratified hydrodynamic turbulence.
These budget equations include production terms caused by the large-scale IGW.
}
\item{We neglect the divergence of the fluxes of TKE, TPE and $F_z$
(i.e., we neglect the divergence of third-order moments).
In the present study, we restrict the analysis to the effects of the
large-scale IGW on the second-order statistics and leave
the third-order moments (the fluxes of energies and
the fluxes of momentum and heat fluxes) for further study.
}
\item{Dissipation rates of TKE, TPE and $F_z$ are expressed using
the Kolmogorov  hypothesis \cite{K41} (see also \cite{MY71}), i.e.,
$\varepsilon_K=E_K/t_{T}$, $\varepsilon_\theta=E_\theta/(C_p \, t_{T})$ and
$\varepsilon^{(F)}=F_z/(C_F \, t_{T})$.
}
\item{We assume that the term $\rho_0^{-1} \, \overline{\langle \theta \,
\nabla_z p \rangle}$ in Eq.~(20) for the vertical turbulent flux of potential
temperature is parameterised by $\tilde C_\theta \, \beta \, \overline{\langle \theta^2
\rangle}$ with $\tilde C_\theta < 1$. This implies that $\beta \, \overline{\langle \theta^2
\rangle} - \rho_0^{-1} \, \overline{\langle \theta \,
\nabla_z p \rangle} = C_\theta \, \beta \, \overline{\langle \theta^2
\rangle}$ with the positive coefficient $C_\theta = 1 -\tilde C_\theta$ that is less than 1.
The justification of this assumption is discussed
in \cite{ZKR07,ZKR13}.
}
\item{The considered stratified hydrodynamic turbulence with large-scale IGW
is shear-free and isotropic in the horizontal plane.
}
\end{itemize}
In the next section we will use Eqs.~(\ref{C1})--(\ref{C4}) to study effects of the large-scale IGW on turbulence.

\section{Effects of large-scale IGW on turbulence: steady-state regime}

In this section we study the effects of large-scale internal gravity waves on turbulence,
using the steady state version of Eqs.~(\ref{C1})--(\ref{C4}). Solving the system of these equations,
we obtain the turbulent kinetic energy $E_K$ and the vertical turbulent flux $F_z$
of the potential temperature:
\begin{eqnarray}
E_K &=& {2 C_F \over 3} \, {(\ell_0 N_0)^2 \over {\rm Ri}_{_{\rm W}} \, \hat W} ,
 \label{D1}\\
F_z &=& -K_H \, {N_0^2 \over \beta} ,
  \label{D2}
\end{eqnarray}
where $N_0=N(z=z_0)$ and the coefficient of turbulent (eddy) diffusivity is
\begin{eqnarray}
K_H &=& \left({2 C_F \over 3}\right)^{3/2} \, {\ell_0^2 \, N_0 \left(\hat W - 1\right) \over \left({\rm Ri}_{_{\rm W}} \, \hat W \right)^{3/2}} .
 \label{D3}
\end{eqnarray}
Here we have introduced the following two key parameters characterizing
effects of the large-scale IGW on turbulence:

(i) the wave Richardson number:
\begin{eqnarray}
{\rm Ri}_{_{\rm W}} = {N_0^2 \, H^2 \over \left(\Gamma(\mu) \, {\rm Pr}_T^{(0)}\right) \, E^{\rm W}} ;
 \label{D4}
\end{eqnarray}

(ii) the parameter $\hat W$:
\begin{eqnarray}
 \hat W \equiv {- \overline{\tau_{ij}^{\rm W} \, \nabla_j V^{\rm W}_i} \over E_K/t_T} = C_\tau \, \Psi(Q,\mu) \, {E^{\rm W} \, \ell_0^2 \over E_K \, H^2} ,
 \label{D5}
\end{eqnarray}
where $E^{\rm W}$ is the total wave energy, $H$ is the hight of the layer where the large-scale IGW propagate,
the dimensionless functions $\Gamma(\mu)$ and $\Psi(Q,\mu)$ are given below, $\mu$
is the exponent of the energy spectrum of the large-scale IGW and $Q=\left[N(z)/N(z_0)\right]^2$ is the dimensionless lapse rate.

The reason for using these parameters for characterization of the effects of large-scale IGW
on turbulence is as follows.
We consider a shear-free turbulence. The turbulence is produced only by large-scale IGW, so that
the classical gradient Richardson number, ${\rm Ri}_g=N^2/S_{\rm wind}^2$ tends to infinity,
because the mean wind shear, $S_{\rm wind}$, vanishes.
The most appropriate parameter in this case is the effective Richardson number
${\rm Ri}_{_{\rm W}}$ associated with the amplitude of the wave.
For instance, the large wave Richardson number implies a low-amplitude wave.

Note that the Froude number ${\rm Fr} = U/(L N_0)$ used in fluid dynamics is
related to the wave Richardson number ${\rm Ri}_{_{\rm W}}$ as ${\rm Fr} = {\rm Ri}_{_{\rm W}}^{-1/2}$,
where the velocity $U = \sqrt{E^W}$ is related to the wave energy $E^{\rm W}$ and the scale
$L = H \, \left(\Gamma(\mu) \, {\rm Pr}_T^{(0)}\right)^{-1/2}$ is proportional
to the hight $H$ of the layer in which the large-scale IGW propagate.
We use the notion of the wave Richardson number following tradition of the atmospheric physics
and meteorology, where different kinds of Richardson numbers (e.g., the gradient Richardson number
and the flux Richardson number) are used.

Another important parameter, $\hat W$, is defined as the ratio
of the TKE production rate caused by the internal gravity waves to the dissipation rate of TKE.
This parameter describes the efficiency of the turbulence production by waves.
It depends on the wave Richardson number, $\hat W=\hat W\left({\rm Ri}_{_{\rm W}}\right)$,
and it can be interpreted as the squared ratio of the wave shear, $\sqrt{E^{\rm W}} / H$,
and the turbulent shear, $\sqrt{E_K} / \ell_0$.
The function $\hat W\left({\rm Ri}_{_{\rm W}}\right)$ is determined by the cubic algebraic equation:
$\hat W^3 + B_1 \hat W^2 + B_2 \hat W + B_3 =0$, with coefficients $B_k$ given in Appendix~\ref{Appendix-B}.

The function $\Gamma(\mu)$ in Eq.~(\ref{D4}) depends on the exponent $\mu$ of the IGW energy spectrum.
In particular, for a power law, $k^{-\mu}$, energy spectrum of IGW existing in the range of the wavenumbers,
$H^{-1} \leq k \leq L_W^{-1}$, the function $\Gamma(\mu)$ is given by: $\Gamma(\mu)=|C_\mu|
(H/L_W)^{3-\mu}$ for $1<\mu<3$; $\Gamma(\mu)=4 \ln(H/L_W)$ for $\mu=3$ and
$\Gamma(\mu)=C_\mu$ for $\mu >3$, where $C_\mu=2(\mu-1)/(\mu-3)$ \cite{ZKR09}.
The function $\Psi(Q,\mu)$ in Eq.~(\ref{D5}) that is related to the parameter $\hat W$,
is given by
\begin{eqnarray}
\Psi(Q,\mu) = {2 \over 3} \left[1 + 3 (Q-1)A_z\right] \Gamma(\mu) .
 \label{G1}
\end{eqnarray}

Now using the steady-state version of Eqs.~(\ref{C1})--(\ref{C4}),
we determine various dimensionless parameters versus the wave Richardson number.
Assuming for simplicity constant Brunt-V\"{a}is\"{a}l\"{a} frequency which implies that $Q=1$,
we obtain the vertical anisotropy parameter, $A_z = E_z/E_K$, and the ratio of turbulent potential and kinetic energies, $E_P/ E_K$:
\begin{eqnarray}
A_z &=& {5 - \hat W \, (1- A_{z}^\ast) \, (1-{\rm Ri}_{\rm f}^\ast) \over 5 + 2 \hat W (1- A_{z}^\ast) \, (1-{\rm Ri}_{\rm f}^\ast)} ,
\label{D7}\\
{E_P\over E_K} &=& C_P \, \left[\hat W\left(1 + 1/{\rm Pr}_{_{T}}^{(0)}\right)-1\right],
 \label{D6}
\end{eqnarray}
where $E_P = (\beta/N)^2 E_\theta$ is the turbulent potential energy, $A_{z}^\ast$ and ${\rm Ri}_{\rm f}^\ast$ are the vertical anisotropy parameter and the flux Richardson number in the limit of very large gradient Richarson number in the absence of the IGW, respectively, and ${\rm Pr}_{_{T}}^{(0)}$ is the turbulent Prandtl number
for a nonstratified turbulence (at zero gradient Richardson number).
Equation~(\ref{D7}) determines the vertical anisotropy parameter $A_{z}$.
Since $A_{x} + A_{y} + A_{z}=1$ and $A_{x} = A_{y}$, we obtain that $A_{x} = A_{y} = (1- A_{z})/2$,
where the horizontal anisotropy parameters are defined as $A_{x} = E_x/E_K$ and $A_{y} = E_y/E_K$.
The turbulent viscosity, $K_M=2 C_\tau A_z E_K^{1/2} \ell_0$, is given by
\begin{eqnarray}
K_M &=& \left({2 C_F \over 3}\right)^{1/2} \, {2 C_\tau \, \ell_0^2 N_0 \over \left[{\rm Ri}_{_{\rm W}} \, \hat W\left({\rm Ri}_{_{\rm W}}\right)\right]^{1/2}} ,
\label{D8}
\end{eqnarray}
so that the turbulent Prandtl number is
\begin{eqnarray}
{\rm Pr}_{_{T}} = {3 C_\tau \over C_F} \, \left[{ {\rm Ri}_{_{\rm W}} \, \hat W\left({\rm Ri}_{_{\rm W}}\right) \over \hat W -1} \right] \left( {C_A - \hat W \over C_A + 2 \hat W} \right) ,
\label{D9}
\end{eqnarray}
where $C_A=5 \, (1- A_{z}^\ast)^{-1} \, (1-{\rm Ri}_{\rm f}^\ast)^{-1}$.
Equations~(\ref{D1})--(\ref{D3}) and (\ref{D6}) allow us to determine the nondimensional ratio $F_z^2/ E_K E_\theta$:
\begin{eqnarray}
{F_z^2 \over E_K E_\theta} &=& {2 C_F \over 3 C_P} \, \left[{(\hat W -1)^2 \over
{\rm Ri}_{_{\rm W}} \, \hat W\left({\rm Ri}_{_{\rm W}}\right)} \right]
\nonumber\\
&& \times \left[\hat W\left(1 + {1 \over {\rm Pr}_{_{T}}^{(0)}}\right)-1\right]^{-1} .
\label{D10}
\end{eqnarray}

Let us discuss the choice of the dimensionless empirical constants in the developed theory. There are two well-known universal constants: the limiting value of the flux Richardson number ${\rm Ri}_{\rm f}^\ast=0.25$ for an extremely strongly stratified turbulence (i.e., at infinite gradient Richardson number)
and the turbulent Prandtl number ${\rm Pr}_{_{T}}^{(0)}=0.8$ for a nonstratified turbulence (at zero gradient Richardson number).
The vertical anisotropy parameter for an extremely strongly stratified turbulence is $A_{z}^\ast =0.03$
and the constants $C_F=C_\tau/{\rm Pr}_{_{T}}^{(0)} = 0.125$, where $C_\tau$ is the coefficient determining the turbulent viscosity  ($K_M=2 C_\tau A_z E_K^{1/2} \ell_0$) for a non-stratified turbulence. The constant $C_P=0.417$ describes the deviation of the dissipation timescale of $E_\theta=\overline{\langle \theta^2 \rangle}/2$ from the dissipation timescale of TKE. The constants $C_F$, $C_P$ and $A_{z}^\ast$ are determined from numerous meteorological observations, laboratory experiments and LES (see details in \cite{ZKR13}).
The results essentially depends on the empirical constant $C_\theta$, e.g.,
the constant $C_\theta=1/15$ is chosen to get a small parameter $A_{z}^\ast$ to reproduce a quasi-two-dimensional turbulence for an extremely strongly stratified turbulence.

\begin{figure}
\centering
\includegraphics[width=9.0cm]{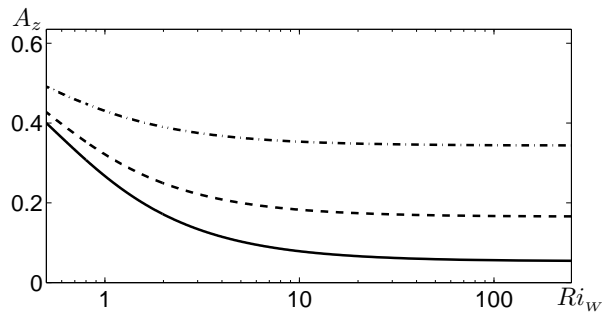}
\caption{\label{Fig1}
The anisotropy parameter $A_z$ versus the wave Richardson number ${\rm Ri}_{_{\rm W}}$
for different values of the parameter $C_\theta$:  $1/15$ (solid), $0.1$ (dashed) and $0.217$  (dashed-dotted).}
\end{figure}

In Fig.~\ref{Fig1} we show the dependence of the vertical anisotropy parameter $A_z= E_z/E_K$ on the wave Richardson number ${\rm Ri}_{_{\rm W}}$ for different values of the parameter $C_\theta$.
The anisotropy parameter depends strongly on the empirical constant $C_\theta$, e.g., it becomes negative, $A_z<0$, when $C_\theta<1/15$.
This indicates that the system does not reach a steady-state.

\begin{figure}
\centering
\includegraphics[width=9.0cm]{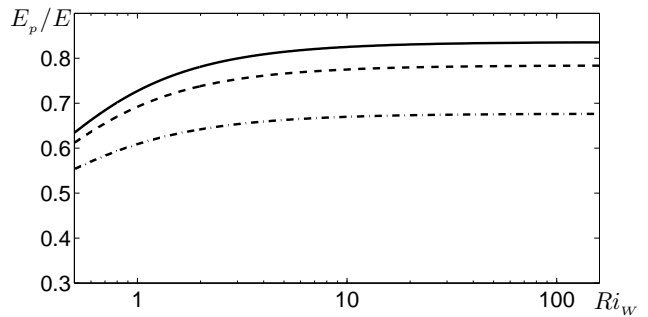}
\caption{\label{Fig2}
The ratio of the turbulent potential energy, $E_P$, to the total turbulent energy $E=E_K+E_P$ versus the wave Richardson number ${\rm Ri}_{_{\rm W}}$ for different values of the parameter $C_\theta=$  $1/15$ (solid), $0.1$ (dashed) and $0.217$  (dashed-dotted).}
\end{figure}

The ratio of the turbulent potential energy, $E_P$, to the total turbulent energy $E=E_K+E_P$ versus the wave Richardson number ${\rm Ri}_{_{\rm W}}$ is shown in Fig.~\ref{Fig2}. It is seen here that for large wave Richardson number (i.e., for the low amplitude IGW), the turbulent potential energy is about 90 \% of total turbulent energy. Without waves, the ratio $E_P/E$ is less than 0.2, while in the presence of the large-scale IGW this ratio can reach 0.9.

\begin{figure}
\centering
\includegraphics[width=8.5cm]{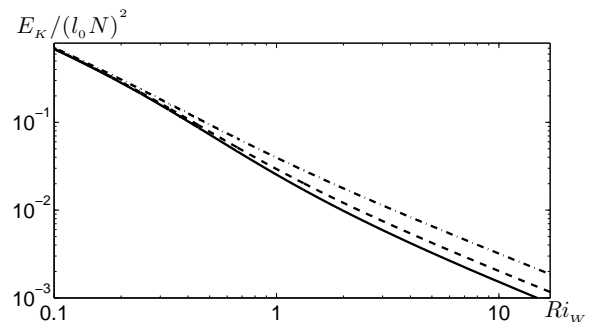}
\caption{\label{Fig3}
The non-dimensional turbulent kinetic energy $E_K/(\ell_0 N_0)^2$ versus ${\rm Ri}_{_{\rm W}}$ for different values of the parameter $C_\theta=$  $1/15$ (solid), $0.1$ (dashed) and $0.217$  (dashed-dotted).}
\end{figure}

The non-dimensional turbulent kinetic energy $E_K/(\ell_0 N_0)^2$ versus ${\rm Ri}_{_{\rm W}}$ [see Eq.~(\ref{D1})] is shown in Fig.~\ref{Fig3}. It is clearly seen in this figure, that the turbulent kinetic energy decreases rapidly with the increase of ${\rm Ri}_{_{\rm W}}$, and $E_K$ is less than the Ozmidov energy scale, $(\ell_0 N_0)^2$.
Note that the Ozmidov length scale, $\sqrt{\varepsilon_K/N_0^3}$,
is well known as a rough threshold between
anisotropic scales and isotropic ones, where
$\varepsilon_K$ is the dissipation rate of TKE.

\begin{figure}
\centering
\includegraphics[width=9.0cm]{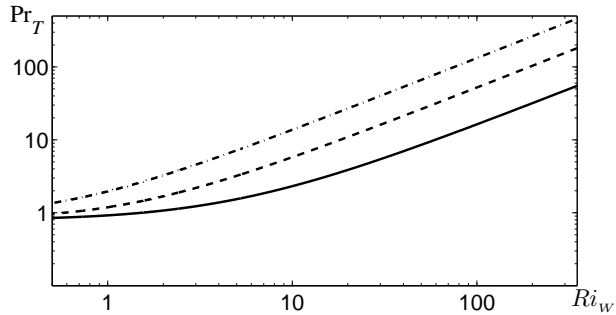}
\caption{\label{Fig4}
The turbulent Prandtl number ${\rm Pr}_{_{T}}$ versus the wave Richardson number ${\rm Ri}_{_{\rm W}}$ for different values of the parameter $C_\theta=$  $1/15$ (solid), $0.1$ (dashed) and $0.217$  (dashed-dotted).}
\end{figure}

\begin{figure}
\centering
\includegraphics[width=8.5cm]{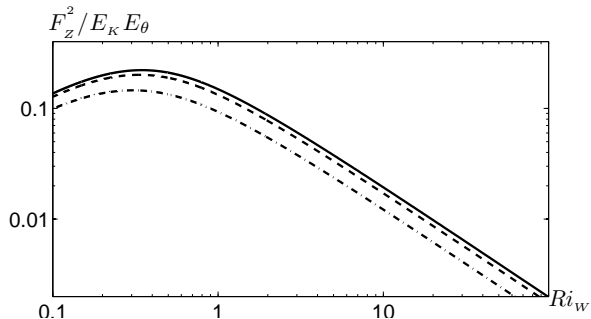}
\caption{\label{Fig5}
The non-dimensional squared potential temperature flux $F_z^2/ E_K E_\theta$ versus the wave Richardson number ${\rm Ri}_{_{\rm W}}$ for different values of the parameter $C_\theta=$  $1/15$ (solid), $0.1$ (dashed) and $0.217$  (dashed-dotted).}
\end{figure}

In Figs.~\ref{Fig4} and~\ref{Fig5} we show the turbulent Prandtl number ${\rm Pr}_{_{T}}$ and the non-dimensional squared potential temperature flux $F_z^2/ E_K E_\theta$ versus the wave Richardson number ${\rm Ri}_{_{\rm W}}$, respectively. It can be seen in these figures that the turbulent Prandtl number increases with decrease of the wave energy $E^{\rm W}$, becomes quite large, and the heat transfer becomes weaker, i.e., the ratio  $F_z^2/ E_K E_\theta$ decreases with decrease of the wave energy.

\section{Two-way coupling between turbulence and large-scale IGW}

In this section we consider the two-way coupling of stably-stratified turbulence and large-scale IGW.
The large-scale IGW emitted at a certain level, propagate
upward, and the losses of wave energy cause the production of turbulence energy.
Equation~(\ref{B10}) for the total wave energy,
$E^{\rm W}$, reads:
\begin{eqnarray}
{\partial E^{\rm W} \over \partial t} + \nabla_z \left(V^{\rm g} \, E^{\rm W}\right) = -\gamma_d E^{\rm W},
 \label{F1}
\end{eqnarray}
where the damping rate, $\gamma_d$, of the large-scale IGW is given by
\begin{eqnarray}
\gamma_d = C_F \, \left(1 + {\rm Pr}_{_{T}}^{(0)}\right) \, \Psi(Q,\mu) \, {\ell_0 \over H^2}
\, E_K^{1/2} ,
 \label{F2}
\end{eqnarray}
$V^{\rm g} = (1-\mu^{-1}) \, N_0 \, H \, f(Q)$ is the group velocity of the large-scale IGW and the function $f(Q)$ is
\begin{eqnarray}
f(Q) = {(Q-1)^{1/2} \over Q} \,\int_0^{\pi/2} \left(1 + {\cos^2 \vartheta \over Q-1}\right)^{1/2} \, \sin^2 \vartheta \,d \vartheta .
\nonumber\\
 \label{F3}
\end{eqnarray}

\begin{figure}
\centering
\includegraphics[width=8.5cm]{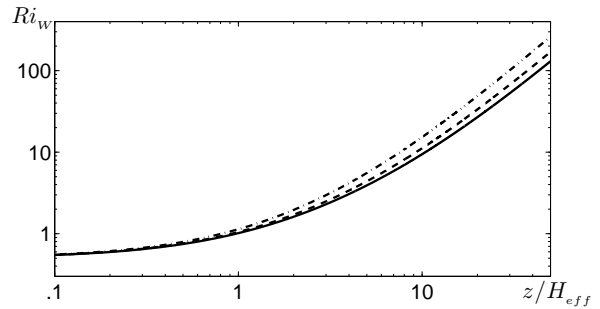}
\caption{\label{Fig6}
The vertical profile of parameter the wave Richardson number ${\rm Ri}_{_{\rm W}}$ for different values of the parameter $C_\theta=$  $1/15$ (solid), $0.1$ (dashed) and $0.217$  (dashed-dotted).}
\end{figure}

The steady-state version of Eq.~(\ref{F1}) can be written in the form of the nonlinear Byger equation, $\nabla_z I = - \kappa \, I$, where $I=V^{\rm g} \, E^{\rm W}$ and $\kappa=\gamma_d/V^{\rm g}$.
This equation can be also rewritten as
\begin{eqnarray}
\nabla_z {\rm Ri}_{_{\rm W}} = {\sqrt{{\rm Ri}_{_{\rm W}}} \over H_{\rm eff} \, \hat W\left({\rm Ri}_{_{\rm W}}\right)}  ,
 \label{F6}
\end{eqnarray}
where the effective damping length scale $H_{\rm eff}$ is defined as
\begin{eqnarray}
H_{\rm eff} = C_\mu H \left({H \over \ell_0}\right)^2 ,
 \label{F7}
\end{eqnarray}
and $C_\mu=2(\mu-1)/(\mu-3)$ depends on the exponent $\mu$ of the energy spectrum of the large-scale IGW.
Equation~(\ref{F6}) allows us to obtain the spatial profile of the wave Richardson number.
In particular, the function ${\rm Ri}_{_{\rm W}}(z)$, as the result of
solution of Eq.~(\ref{F6}) is shown in Fig.~\ref{Fig6}.
As follows from Eq.~(\ref{F7}) and Fig.~\ref{Fig6}, the effective damping length scale $H_{\rm eff}$ is much larger than the equilibrium height $H$. This implies that the large-scale IGW penetrate almost the entire atmosphere height and generate weak turbulence. These waves are significantly attenuated at $z = 10 H_{\rm eff}$.

Equation for the total turbulent energy $E=E_K + E_P$ is
\begin{eqnarray}
{\partial E \over \partial t} + \nabla_z \Phi = \Pi - \varepsilon,
 \label{F4}
\end{eqnarray}
where $\Pi = \gamma_d E^{\rm W}$ is the production rate of the total turbulent energy $E$ caused by the damping of the large-scale IGW, and $\varepsilon=E/(C_P t_T)$ is the dissipation rate of the total turbulent energy. Equations~(\ref{F1}) and~(\ref{F4}) yield the budget equation for the sum, $E+E^{\rm W}$ of the turbulent total energy and the wave total energy:
\begin{eqnarray}
{\partial \left(E + E^{\rm W}\right) \over \partial t} + \nabla_z \left(\Phi + V^{\rm g} \, E^{\rm W}\right)= - {E \over C_P t_T} .
 \label{F5}
\end{eqnarray}
This equation describes energy exchange between turbulence and large-scale internal gravity waves.

\begin{figure}
\centering
\includegraphics[width=8.5cm]{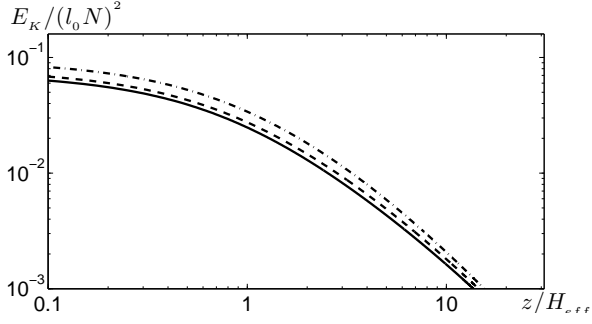}
\caption{\label{Fig7}
The vertical profile of the normalized turbulent kinetic energy, $E_K/(\ell_0 N_0)^2$, for different values of the parameter $C_\theta=$  $1/15$ (solid), $0.1$ (dashed) and $0.217$  (dashed-dotted).}
\end{figure}

\begin{figure}
\centering
\includegraphics[width=8.5cm]{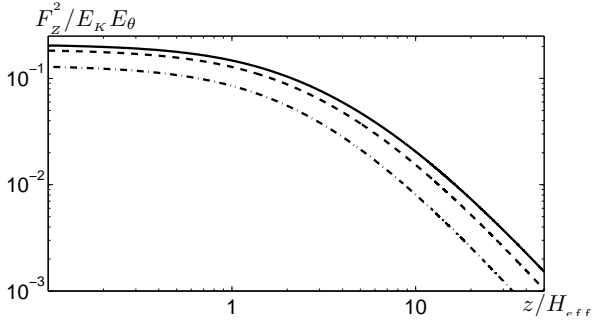}
\caption{\label{Fig8}
The vertical profile of the normalized squared potential temperature flux $F_z^2/ E_K E_\theta$, for different values of the parameter $C_\theta=$  $1/15$ (solid), $0.1$ (dashed) and $0.217$  (dashed-dotted).}
\end{figure}

Let us analyze the two-way interaction between large-scale IGW and stably-stratified turbulence.
In Fig.~\ref{Fig7} we show the vertical profile of the normalised turbulent kinetic energy, $E_K(z)/(\ell_0 N_0)^2$. It decreases rapidly with height reaching the value $2 \times 10^{-3}$ at $z = 10 H_{\rm eff}$.
The vertical profile of the normalized squared turbulent flux $F_z^2/ E_K E_\theta$ of potential temperature (see Fig.~\ref{Fig8}) is similar to the vertical profile of the turbulent kinetic energy, shown in Fig.~\ref{Fig7}.

\begin{figure}
\centering
\includegraphics[width=8.5cm]{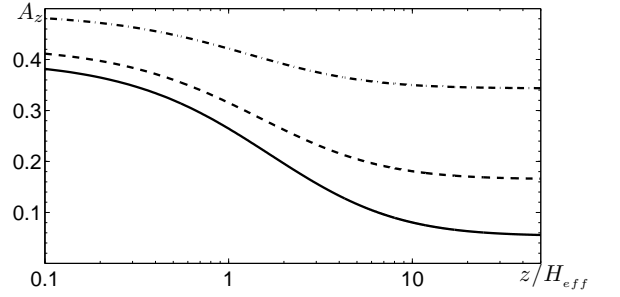}
\caption{\label{Fig9}
The vertical profile of the anisotropy parameter $A_z$, for different values of the parameter $C_\theta=$  $1/15$ (solid), $0.1$ (dashed) and $0.217$  (dashed-dotted).}
\end{figure}

The vertical profile of $A_z$ is shown in Fig.~\ref{Fig9}, which demonstrates that $A_z$ decreases with height.
This implies that anisotropy of turbulence increases with height, and this effect is more significant for $C_\theta=$  $1/15$. For small wave Richardson numbers (i.e., for intensive waves), turbulence is almost isotropic, but for large heights the turbulence anisotropy becomes more significant compared to the case of a sheared stably stratified turbulence.
In particular, when hight $z$ varies from $z= 0.1 \, H_{\rm eff}$ to $z = 50 H_{\rm eff}$, the parameter $A_{z}$ decreases in 10 times (see Fig.~\ref{Fig9}).
The latter implies formation of a ''pancake" structure in turbulent velocity field for large $z$ (see, e.g., \cite{C01,SC18}).

\begin{figure}
\centering
\includegraphics[width=8.5cm]{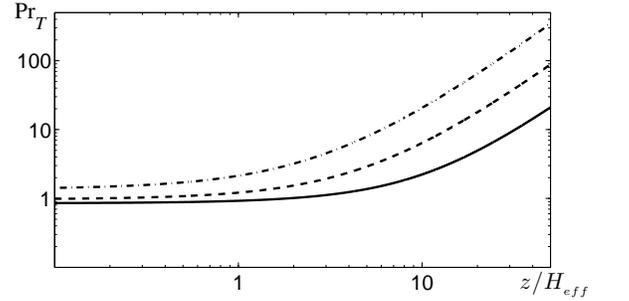}
\caption{\label{Fig10}
The vertical profile of the turbulent Prandtl number, for different values of the parameter $C_\theta=$  $1/15$ (solid), $0.1$ (dashed) and $0.217$  (dashed-dotted).}
\end{figure}

\begin{figure}
\centering
\includegraphics[width=8.5cm]{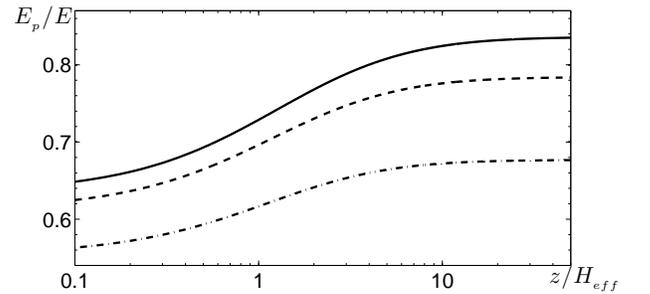}
\caption{\label{Fig11}
The vertical profile of $E_P/E$, for different values of the parameter $C_\theta=$  $1/15$ (solid), $0.1$ (dashed) and $0.217$  (dashed-dotted).}
\end{figure}

The turbulent Prandtl number and the ratio of potential to the total energy $E_P/E$ shown in Figs.~\ref{Fig10} and~\ref{Fig11},
increase with height as the wave becomes less intensive.
It is seen in Fig.~\ref{Fig11} that for the upper layers, the turbulent potential energy is larger than the  turbulent kinetic energy. This implies that the temperature fluctuations dominate over the velocity fluctuations.
In particular, the ratio $E_p/E$ increases up to 0.85 at $z=50 H_{\rm eff}$.
For comparison, in a shear-produced stably stratified turbulence without IGW, the ratio of turbulent potential energy to total turbulent energy, $E_p/E$, reaches 0.15 at very large gradient Richardson numbers (see Fig.~7 in \cite{ZKR13}).
On the other hand, in a shear-produced stably stratified turbulence with IGW (where only the one-way
coupling is taking into account), the maximum ratio $E_p/E$
increases up to 0.45 at very large gradient Richardson numbers
(see Fig.~5 in \cite{ZKR09}).

\section{Conclusions}

\begin{itemize}
\item{Within the new EFB turbulence closure model a system of equations describing two-way interactions between internal gravity waves (IGW) and turbulence has been derived. This system includes the budget equation for the total (kinetic plus potential) energy of IGW, and the equations for the kinetic and potential energies of turbulence, turbulent heat fluxes for waves and flow for arbitrary stratification. The general physical picture in two-way coupling between IGW and turbulence is following: waves emitted at a certain level (in case if there is no refraction), propagate upward. The losses of wave energy cause the production of turbulence. Therefore, waves transfer energy, while turbulence causes its losses. We have shown that more intensive waves penetrate in a shorter distance, while less intense IGW penetrate in larger distances. This is caused by the nonlinear effects, where the more intense IGW produce more strong turbulence that results in more intensive damping of IGW.}
\item{The analysis of the effects of IGW on the anisotropy of turbulence has shown that for less intensive waves the turbulence anisotropy is stronger. Low amplitude waves produce anisotropic turbulence with a low energy, and the total turbulent energy consists up to 90 \% of potential energy.  This property resembles one observed in high altitude tropospheric nearly two-dimensional turbulence.}
\item{We also have demonstrated that the kinetic energy of turbulent fluctuations has the Ozmidov energy scale
    (which is the product of the Brunt-V\"{a}is\"{a}l\"{a} frequency and turbulence integral scale). When turbulence is produced only by waves, the gradient Richardson number tends to infinity, because it is a shear-free turbulence (for which the wave shear is much larger than the wind shear). The most appropriate parameter in this case is the effective Richardson number associated with the amplitude of the wave.
    }
 \end{itemize}

\begin{acknowledgements}
The detailed comments on our manuscript by the two anonymous referees
which essentially improved the presentation of our results,
are very much appreciated.
NK and IR acknowledge support from
the Israel Science Foundation governed by the
Israel Academy of Sciences (grant No. 1210/15).
IR also acknowledges the Research Council of Norway under the FRINATEK (grant  No. 231444)
and the National Science Foundation (grant No. NSF PHY-1748958).
IS and YT acknowledge support from Russian Foundation for Basic Research (RFBR grants
No. 18-05-00292, No. 18-05- 00265, No. 17-05-41117, No. 19-05-00366).
SZ acknowledges support from the Academy of Finland (grant ClimEco No.
314 798/799, FMI-FI), Russian Science Foundation (RSF grant No. 15-17-20009,
IPFAN-RU), Russian Foundation for Basic Research (RFBR grants No. 18-05-60299,
IPFAN-RU, and No. 18-05-60126, INM/MSU-RU), Russian Science Foundation
(RSF grant No. 14-17-00131, Tyumen State University-RU).
NK and IR acknowledge the hospitality of Nordita and
Institute for Atmospheric and Earth System Research (INAR) of
University of Helsinki. IR also acknowledges the hospitality of
the Kavli Institute for Theoretical Physics in Santa Barbara.\\
\end{acknowledgements}

\appendix

\section{\bf Derivation of Eq.~(\ref{L1})}
\label{Appendix-A}

Equation~(\ref{L1}) follows from Eq.~(\ref{B11}). In particular,
substituting Eqs.~(\ref{B3})--(\ref{B4}) and~(\ref{C5})--(\ref{C6}) into Eq.~(\ref{B11}),
performing averaging, and taking into account that for the linear IGW
$E_K^{\rm W} = E_P^{\rm W} = (1/2) E^{\rm W}$,
we arrive at expression~(\ref{L1}) for the dissipation rate $D^{\rm W}$
of the total wave energy $E^{\rm W}$.

A simple and approximate derivation of Eq.~(\ref{L1}) is as follows.
The dissipation rate of the wave kinetic energy $E_K^{\rm W}$ is
$D^{\rm W}_K = - K_M \overline{{\bm V}^{\rm W} \cdot \Delta {\bm V}^{\rm W}}$,
which in ${\bm k}$ space reads $2 K_M k^2 E_K^{\rm W}$.
The dissipation rate of the wave potential energy $E_P^{\rm W}$ is
$D^{\rm W}_P= - (\beta^2/N^2) K_H \overline{\Theta^{\rm W} \, \Delta \Theta^{\rm W}}$,
which in ${\bm k}$ space reads $2 K_H k^2 E_P^{\rm W}$.
Taking into account that for the linear IGW,
$E_K^{\rm W} = E_P^{\rm W} = (1/2) E^{\rm W}$,
we arrive at the expression~(\ref{L1}) for the dissipation rate $D^{\rm W}=D^{\rm W}_K+D^{\rm W}_P$
of the total wave energy $E^{\rm W}$.

\section{\bf The function $\hat W\left({\rm Ri}_{_{\rm W}}\right)$}
\label{Appendix-B}

The function $\hat W\left({\rm Ri}_{_{\rm W}}\right)$ is determined by the following cubic algebraic equation:
$\hat W^3 + B_1 \hat W^2 + B_2 \hat W + B_3 =0$, where
\begin{widetext}
\begin{eqnarray}
B_1 &=& - \biggl[1 - 2 C_\theta C_P  + C_A \biggl(C_\theta C_P \, \left(1 + 1/{\rm Pr}_{_{T}}^{(0)}\right) - {C_F \over 5}\biggr)
+ {2 \over 3{\rm Ri}_{_{\rm W}}} \biggr] \, \left[ C_F - 2 C_\theta C_P \, \left(1 + 1/{\rm Pr}_{_{T}}^{(0)} \right)\right]^{-1},
 \label{T100}\\
B_2 &=& \left[C_A \left(1 + C_\theta C_P\right) + {2 - C_A \over 3 {\rm Ri}_{_{\rm W}} } \right]
\left[ C_F - 2 C_\theta C_P \, \left(1 + 1/{\rm Pr}_{_{T}}^{(0)}\right)\right]^{-1},
 \label{T101} \\
B_3 &=& {C_A \over 3 {\rm Ri}_{_{\rm W}} } \,
\left[C_F - 2 C_\theta C_P \, \left(1 + 1/{\rm Pr}_{_{T}}^{(0)} \right)\right]^{-1} ,
 \label{T102}
\end{eqnarray}
\end{widetext}
\noindent
and $C_A= 5 (1- A_{z}^\ast)^{-1}(1-{\rm Ri}_{\rm f}^\ast)^{-1}$.

\end{document}